\documentclass[sigconf]{acmart}
\usepackage{epsfig}
\usepackage{graphicx}
\usepackage{amsmath}
\usepackage{mathrsfs}
\usepackage{makecell}
\usepackage{booktabs}
\usepackage{multirow}
\usepackage{caption}
\usepackage{algorithmicx}  
\usepackage{algorithm}
\usepackage{algpseudocode}  
\usepackage{enumitem}
\usepackage{subfig}
\usepackage{bm}
\usepackage{colortbl}
\usepackage{arydshln} 

\AtBeginDocument{%
  \providecommand\BibTeX{{%
    \normalfont B\kern-0.5em{\scshape i\kern-0.25em b}\kern-0.8em\TeX}}}

\copyrightyear{2023} 
\acmYear{2023} 
\setcopyright{acmlicensed}\acmConference[SIGIR '23]{Proceedings of the 46th
International ACM SIGIR Conference on Research and Development in
Information Retrieval}{July 23--27, 2023}{Taipei, Taiwan}
\acmBooktitle{Proceedings of the 46th International ACM SIGIR Conference on
Research and Development in Information Retrieval (SIGIR '23), July 23--27,
2023, Taipei, Taiwan}
\acmPrice{15.00}
\acmDOI{10.1145/3539618.3592078}
\acmISBN{978-1-4503-9408-6/23/07}

\begin{document}

\title[Uncertainty-aware Consistency Learning for Cold-Start Item Recommendation]{Uncertainty-aware Consistency Learning for \\Cold-Start Item Recommendation}

\author{Taichi Liu}
\email{liutq20@mails.tsinghua.edu.cn}
\affiliation{%
  \institution{Tsinghua University}
}

\author{Chen Gao}\authornote{Chen Gao is the Corresponding Author.}
\email{chgao96@gmail.com}
\affiliation{%
  \institution{Tsinghua University}
}
\affiliation{%
  \institution{Huawei Noah's Ark Lab}
}

\author{Zhenyu Wang}
\email{wangzy20@mails.tsinghua.edu.cn}
\affiliation{%
  \institution{Tsinghua University}
}

\author{Dong Li}
\email{lidong106@huawei.com}
\affiliation{%
  \institution{Huawei Noah Ark's Lab}
}

\author{Jianye Hao}
\email{haojianye@126.com}
\affiliation{%
  \institution{Huawei Noah's Ark Lab}
}

\author{Depeng Jin}
\email{jindp@tsinghua.edu.cn}
\affiliation{%
  \institution{Tsinghua University}
}

\author{Yong Li}
\email{liyong07@tsinghua.edu.cn}
\affiliation{%
  \institution{Tsinghua University}
}

% Self-training Graph Neural Networks for Long-tail Recommendation

\begin{abstract}
Graph Neural Network (GNN)-based models have become the mainstream approach for recommender systems. Despite the effectiveness, they are still suffering from the cold-start problem, \textit{i.e.}, recommend for few-interaction items. Existing GNN-based recommendation models to address the cold-start problem mainly focus on utilizing auxiliary features of users and items, leaving the user-item interactions under-utilized. However, embeddings distributions of cold and warm items are still largely different, since cold items' embeddings are learned from lower-popularity interactions, while warm items' embeddings are from higher-popularity interactions. Thus, there is a \textit{seesaw phenomenon}, where the recommendation performance for the cold and warm items cannot be improved simultaneously. To this end, we proposed a \textbf{U}ncertainty-aware \textbf{C}onsistency learning framework for \textbf{C}old-start item recommendation (shorten as UCC) solely based on user-item interactions. Under this framework, we train the teacher model (generator) and student model (recommender) with consistency learning, to ensure the cold items with additionally generated low-uncertainty interactions can have similar distribution with the warm items. Therefore, the proposed framework improves the recommendation of cold and warm items at the same time, without hurting any one of them. Extensive experiments on benchmark datasets demonstrate that our proposed method significantly outperforms state-of-the-art methods on both warm and cold items, with an average performance improvement of 27.6\%.

\end{abstract}



\maketitle
\vspace{-0.25cm}
\section{Introduction}\label{sec::intro}
Recommender systems have become a fundamental service for filtering information by learning from collected user behavioral data and inferring users' personalized demands. Currently, GNN-based models, with their strengths in learning from structured data and capturing high-order similarity, have become state-of-the-art approaches in various recommendation tasks~\cite{gao2023survey}. GNNs have a strong ability to incorporate information from the neighboring nodes, helping to understand the local context of nodes. However, it leads to the poor performance for the cold-start recommendation~\cite{volkovs2017dropoutnet, zhang2019star, hu2019hers, zhao2022investigating, abdollahpouri2017controlling,canamares2018should,yin2012challenging}, which occurs when a new item node is added to the graph.
Specifically, there are not enough neighbors for the new item due to few user-item interactions. The cold-start recommendation problem can be particularly challenging in scenarios where the user-item graph rapidly evolves, such as short-video recommendation~\cite{ling2022slapping}. 
% In these cases, the addition of new nodes and edges can significantly affect the distribution between warm items and cold items on the graph, making it difficult for GNNs to generalize to new nodes (cold items) when their neighbors are unknown~\cite{zhao2022investigating, abdollahpouri2017controlling,canamares2018should,yin2012challenging}.

To address this challenge, existing models~\cite{volkovs2017dropoutnet,zhu2021learning, du2022socially, chen2022generative, togashi2021alleviating} have proposed various techniques for incorporating additional information into GNNs, such as social networks, node attributes, and other metadata, which can partly improve the performance for cold-start items.
% These techniques can partly improve the performance of the cold-start problem by using additional context to infer the properties of new nodes.
% without access to their neighbors. 
However, cold items' embeddings are still continuously learned from fewer interactions, while warm items' embeddings are still mainly learned from adequate interactions~\cite{abdollahpouri2019managing, abdollahpouri2021user}, which leads to the issue that the distributions of embeddings of cold and warm items are significantly different. 
% Consequently, the embeddings of cold items cannot reflect the real preferences between user and item, 
That is, the collected interaction of cold items cannot reflect the real item characteristics due to less exposure, thus resulting in the \textit{seesaw phenomenon}, \textit{i.e.}, improves the recommendation of either the cold or warm items will hurt the other~\cite{chen2020bias, zhang2021causal}.
% particularly in GNNs where the interactions (edges) play a critical role.

% \textbf{popularity bias} remains a persistent issue in supervised training processes that rely on user-item interactions, owing to the fact that frequently interacted warm items with higher popularity can skew the recommendations towards popular items~\cite{abdollahpouri2019managing, abdollahpouri2021user}. Even if these models successfully generate the embeddings of cold items, the distribution of embeddings for cold and warm items is different due to popularity bias. This can lead to the seesaw phenomenon, which improves the recommendation of either the cold or warm items, but hurts the other.

To this end, we proposed a framework for bridging the distribution gap between cold items and warm items in the cold-start recommendation, with two key designs included: a) \textbf{Uncertainty-aware Interaction Generation}, b) \textbf{Graph-based Teacher-Student Consistency Learning}. 
In particular, we first introduce low-uncertainty interactions and take use of generated interactions for embedding learning, to address the distribution difference which causes the seesaw phenomenon. 
We then develop a teacher-student training paradigm that improves the robustness~\cite{park2006naive,wei2021contrastive} in embedding learning for cold and warm items. 
Specifically, we propose a contrastive loss to keep the consistency of item embeddings before and after the interaction generation. 
We further fine-tune the student model on the recommendation task, 
maintaining the consistency between the teacher model and student model. 
Our contributions can be summarized as follows:

\begin{itemize}[leftmargin=*]
   \item We propose approach the cold-start problem from the novel perspective of addressing the distribution difference between cold and warm items, by proposing a general UCC framework.
   % and improves the recommendation for both the warm and cold items.
   %in which the teacher model generates low-uncertainty interactions for the student model to modulate representations in similar distribution for cold and warm items.
   \item Under the framework, we first address the distribution gap between cold and warm items through our uncertainty-aware interaction generation, and keep both item-level and model-level similarity with consistency learning, which can improve model's robustness.
    \item We conduct extensive experiments to  demonstrate the effectiveness of our framework. The results shown an average improvement of 27.6\% in ranking metrics compared to the baseline method on two widely-used benchmark datasets.
\end{itemize}

% \vspace{-4mm}
% \input{1to2.preli.tex}
% \section{Methodology}\label{sec::method}
%\chenc{if we talk about debiasing, we should discuss about debiasing papers; please cite the survey}

\begin{figure*}[t]
\includegraphics[width=\textwidth]{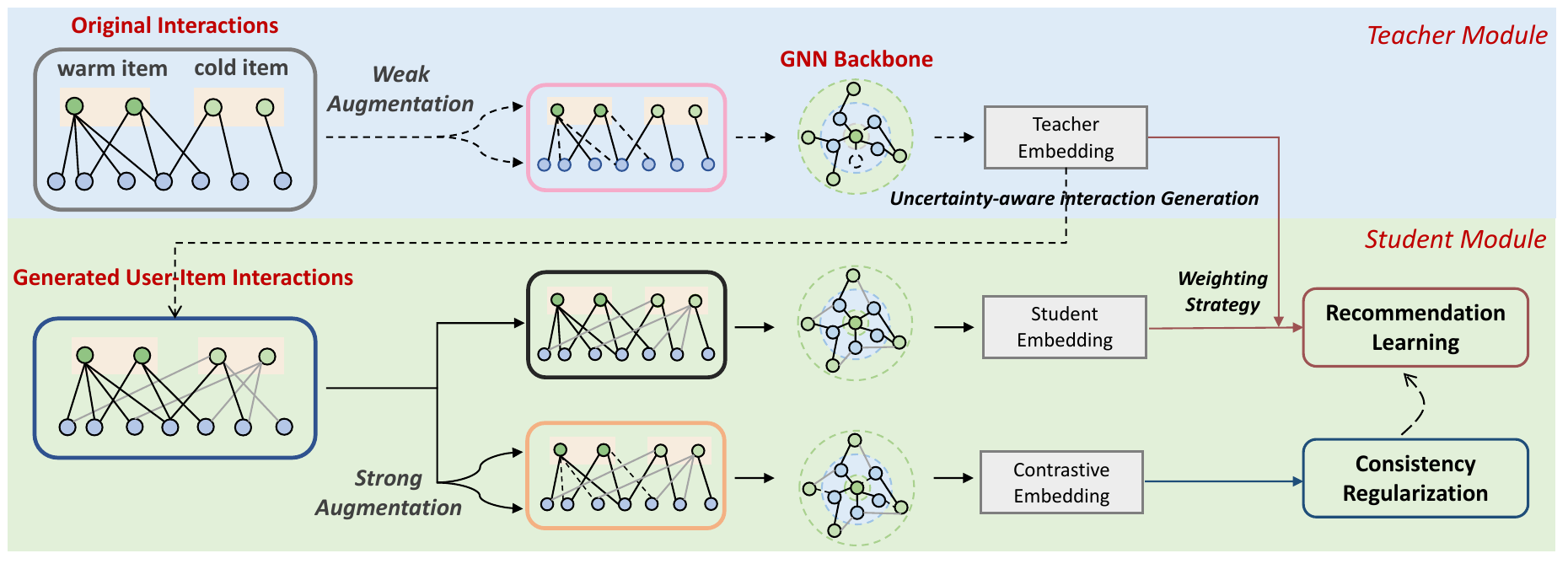}
% \caption{Our proposed approach}
\caption{The overall framework of UCC. A teacher model (the upper blue frame) is first trained under weak augmentation regularization. Then interactions are generated with uncertainty-aware selection process. Finally the generated interactions serve for the student model training (the lower green frame). Strong augmentation based consistency regularization and integrated teacher representation increase the robustness of the student model.}
\label{fig:model}
\end{figure*}

\section{Problem Formulation}
%\chenc{we should emphasize long-tail recommendation here;}
\noindent\textbf{Notations.} The common GNN-based model paradigm can be formulated as follows: Let $\mathcal{U}=\{u\}$, $\mathcal{I}=\{i\}$ denote the set of users and items respectively. The interactions of user and item can be regard as $\mathcal{O}=\left\{(u, i^{+}) \mid u \in \mathcal{U}, i^{+} \in \mathcal{I}\right\}$, where each pair represents each observed feedback.

\noindent\textbf{Input:} Most GNN-based models treat user-item interactions as a bipartite graph, where the node set is $\mathcal{V} = \mathcal{U} \cup \mathcal{I}$ and the edge set is $\mathcal{G} = (\mathcal{V}, \mathcal{O}^{+})$. In the training process of graph representation learning, we use $\textbf{E}_u = [e_{u_{1}},\cdots,e_{u_{M}}]\in \mathbb{R}^{M \times D}, \textbf{E}_i = [e_{i_{1}},\cdots,e_{i_{N}}]\in \mathbb{R}^{N \times D}$ 
to denote the user embedding and item embedding, where D is the dimension size of embedding, $M, N$ are the number of users and items respectively.

\noindent\textbf{Output:} Recommender models estimate the relations of unobserved user-item pairs through the dot products of their embeddings ~\cite{grover2016node2vec,he2017neural,hsieh2017collaborative}. Specifically, the scores $\{s_{mn}\}$ of a given user m and a given item n, where $s_{mn}$ is calculated as:
\begin{equation}
    s_{mn} = e_{u_{m}}e_{i_{n}}^T,
\end{equation}
then rank items by their scores - a high score means that the user prefers. The top-k items from the ranking list are adopted into a candidate set and recommended to the user. 
% The recall process for each user can be formulated as:
% \begin{equation}
%     \widehat{\mathcal{I}}_\textbf{candidate}= \text{Top-K}(S_m)
% \end{equation}

\section{Methodology}
\subsection{Uncertainty-aware Interaction Generation}

For the cold-start recommendation, one main reason for the limited performance is the different distribution between cold items and warm items caused by popularity bias. Hence, we propose to generate similar distributed interactions for cold items as warm items, and we need to select accurate and reliable interactions from unobserved interactions.

However, since totally unbiased and accurate interactions for the original data used are unavailable, especially for cold items, it is hard to discriminate whether an interaction is reliable. We thus propose to estimate the uncertainty degree of each user-item interaction, and select low-uncertainty interactions for each item. We estimate the uncertainty of the particular interaction between the user $u_m$ and item $i_n$ by their similarity. We apply the cosine distance to calculate the similarity:
\begin{equation}
d_{m n}=\frac{\left|e_{u_{m}}e_{i_{n}}^T\right|}{\left\|e_{u_{m}}\right\|\left\|e_{i_{n}}\right\|},
\end{equation} 

With the pre-trained recommender, we have the ranking scores $\{s_{mn}\}_{n=1}^N$ of each item $i_n$  for all users. The average of all ranking scores measures the overall uncertainty degree of the item $i_n$ as follows:
\begin{equation}
    \bar{s}_n = \frac{1}{M}\sum\limits_{m=1}^{M}s_{mn},
\end{equation}
%The ranking score is a reasonable and simple representation of how certain the model is about the specific interaction.
where $\bar{s}_n$ provides a description of the overall interaction uncertainty of the item $i_n$. We aim to force the cold items with generated interactions to have a similar distribution as the warm items to improve the recommendation performance for both cold and warm items. A small $d_{nm}$ interactions are regarded as uncertain ones and are filtered out in the generation phase. This selection strategy based on the interaction uncertainty guides the generator on more certain interactions to mitigate the popularity bias in the cold-start problem. The selection is as follows:
\begin{equation}
\widehat{\mathcal{O}}_n= I(d_{m n} > \alpha \bar{s}_n),
\end{equation}
where $\alpha$ is a pre-defined parameter and $I$ is the indicator function.

\subsection{Teacher-Student Consistency Learning}
The consistency learning for our framework consists of two parts: 
\subsubsection{Item-level consistency learning.} We adopt the contrastive loss for the item embeddings between pre-generation and post-generation. Inspired by~\cite{sohn2020fixmatch}, we adopt two augmentations - strong augmentation and weak augmentation.
Specifically, the weak augmentation drops out the edges in the graph with a dropout ratio $\rho$, which can be formulated as follows:
\begin{equation}
    \mathcal{G}^w = (\mathcal{V}, \mathcal{M} \cdot \mathcal{O}^{+}),
\end{equation}
where $\mathcal{M} \in \{0, 1\}^{|\mathcal{O}^{+}|}$ is the masking vector.
The strong augmentation based on generated labels adds more edges to the graph, which can be formulated as follows,
\begin{equation}
    \mathcal{G}^s = (\mathcal{V}, \mathcal{O}^{+} + \widehat{\mathcal{O}}),
\end{equation}

Two augmentation operations on the graph generates two different views of each node. 
For the item node $i$, we denote its two different views as $z_i'$, $z_i''$ (on weak and strong graphs respectively). 
We perform a consistency regularization for these different views, to encourage the similarity between different views of the same node.
To implement the consistency regularization, the contrastive loss \cite{chen2020simple} is as follows,
\begin{equation}
\begin{aligned}
\mathcal{L}_\textbf{cr, item} = \sum_{i \in \mathcal{I}}-\log \frac{\exp ({\rm sim} (z_i', z_i'') / \tau )}{\sum_{j \in \mathcal{I}} \exp ({\rm sim}(z_i', z_j'') / \tau)},
\end{aligned}
\end{equation}
where $\mathcal{L}_\textbf{cr, item}$ denotes the item-side consistency regularization loss for the teacher and the student model, respectively.  Here ${\rm sim}(\cdot, \cdot)$ denotes the similarity function, for which we use cosine similarity, and $\tau$ is a pre-defined hyper-parameter. The user-side consistency regularization loss $\mathcal{L}_\textbf{cr, user}$ can be similarly calculated, and we omit it due to the limit of space. The final consistency loss $\mathcal{L}_\textbf{cr} = \mathcal{L}_\textbf{cr, item} + \mathcal{L}_\textbf{cr, user}$ serves for consistency regularization. For the recommendation loss we use Bayesian personalized ranking (BPR)~\cite{rendle2012bpr} loss to optimize the representation learning, which can be formulated as follows,
\begin{equation}
\mathcal{L}_\textbf{rec}=\sum_{(u, i^{+}, i^{-}) \in O}-\ln \sigma\left(\hat{y}_{u i^{+}}-\hat{y}_{u i^{-}}\right)+\lambda\|\Theta\|_{2}^{2},
\end{equation}
where $(u,i^{+})$,  $(u, i^{-})$ represent observed/unobserved interaction pairs and $\|\Theta\|_{2}^{2}$ is
L2-regularization of model's parameters. Then our final loss of the whole framework is:
\begin{equation}
\mathcal{L}_\textbf{total}= \mathcal{L}_\textbf{rec} + \mu \mathcal{L}_\textbf{cr},
\end{equation}
where $\mu$ is a hyper-parameter.

\subsubsection{Model-level consistency learning.}
To maintain the consistency between the teacher model and student model, We denote the embedding of the student model's embedding as $\textbf{E}^s$. After each iteration, we propose to accumulate the teacher embedding into the student embedding:
\begin{equation}
\textbf{E}^s \leftarrow \gamma \textbf{E}^s + (1-\gamma) \textbf{E}^t,
\end{equation}
where $\gamma$ denotes a momentum term that controls the impact of the teacher model. Here the accumulation step includes both user embeddings and item embeddings.

\section{Experiments}\label{sec::experiments}

% To justify the superiority of our framework and reveal the reasons for its effectiveness, we conduct extensive experiments and answer the following research questions (RQs):

% $\bullet$ \textbf{RQ1:} How does our method perform w.r.t. top-K recommendation compared to the state-of-the-art models?

% $\bullet$ \textbf{RQ2:} How does uncertainty-aware interaction generation alleviate the distribution problem of cold and warm interactions?

% $\bullet$ \textbf{RQ3:} How does teacher-student consistency learning comprehensively improve recommendation?

\subsection{Experimental Setting}

\noindent \textit{\textbf{Datasets.}} We conduct experiments on benchmark datasets: Yelp and Amazon-Book, following the same 10-core setting as \cite{he2017neural, he2020lightgcn}.
We split all user-item interactions into training, validation, and testing set with the ratio of 7:1:2, evaluating the top-$K$ recommendation performance with two widely-used metrics Recall@K and NDCG@K where $K = 20$, following \cite{he2020lightgcn}.
% \begin{table}[h]
% \caption{Statistics of the datasets.}
% \begin{tabular}{c|c|c|c|c}
% \hline 
% \bf Dataset & \bf \#Users & \bf \#Items & \bf \#Interactions & \bf Density\\
% \hline \hline 
% Yelp & 31,668 & 38,048 & 1,561,406 & 0.00130 \\
% Amazon-Book & 52,643 & 91,599 & 2,984,108 & 0.00062\\
% \hline
% \end{tabular}
% \label{tab:dataset}
% \end{table}\\

\noindent \textit{\textbf{Baselines.}} We select LightGCN\cite{he2020lightgcn} as the backbone of our GNN-based model. Since we mainly focus on modeling user-item interactions but not the features of items, we compare our model with state-of-the-art recommendation models which can be classified into two classes: generative model NB\cite{saito2020asymmetric} and denosing model IRBPR\cite{wang2021implicit}, ADT\cite{wang2021denoising}, SGL\cite{wu2021self} for user-item interactions.

\begin{table}[t]
\caption{Performance comparison on benchmark datasets.}
%\resizebox{\columnwidth}{!}{
\begin{tabular}{c|cc|cc}
\hline
\bf Dataset & \multicolumn{2}{c|}{ \bf Yelp } & \multicolumn{2}{c}{ \bf Amazon-Book }  \\ \hline
\bf Metric &\bf  Recall &\bf  NDCG & \bf Recall &\bf  NDCG \\
\hline 
LightGCN  & 0.0639 & 0.0515  & 0.0410 & 0.0318\\
IRBPR & 0.0538 & 0.0438 & 0.0325 & 0.0246 \\
NB & 0.0640 & 0.0526 &  0.0425 & 0.0328\\
ADT &  \underline{0.0691} & 0.0388 & 0.0335 & 0.0176\\
SGL & 0.0673 & \underline{0.0554} & \underline{0.0477} & \underline{0.0378}\\
UCC & \textbf{0.0713} & \textbf{0.0587} & \textbf{0.0523} & \textbf{0.0410}\\
\hline
\hline
Improv. & \textbf{+11.6\%} & \textbf{+14.0\%} & \textbf{+27.6\%} & \textbf{+28.9\%}  \\
\hline
\end{tabular}
%}
\vspace{-0.3cm}
\label{tab:mainrr}
\end{table}
%\subsubsection{Implementation Details}

%All models in our experiment section are trained from scratch, initialized with the Xavier method \cite{glorot2010understanding}. We adopt the Adam optimizer with learning rate of 0.001 and the batch size of 2048, following existing works~\cite{XX}.
%The early stopping strategy is the same as NGCFand LightGCN. 
%For the hyper-parameter setting, we utilize L3 norm for embedding normalization and adopt the top five users for our item-side pseudo labels. The weight value\chenc{XXX}, $\gamma$, is set to 0.3.

\begin{figure}
   \centering
\subfloat[Yelp]{\includegraphics[width=0.49\columnwidth]{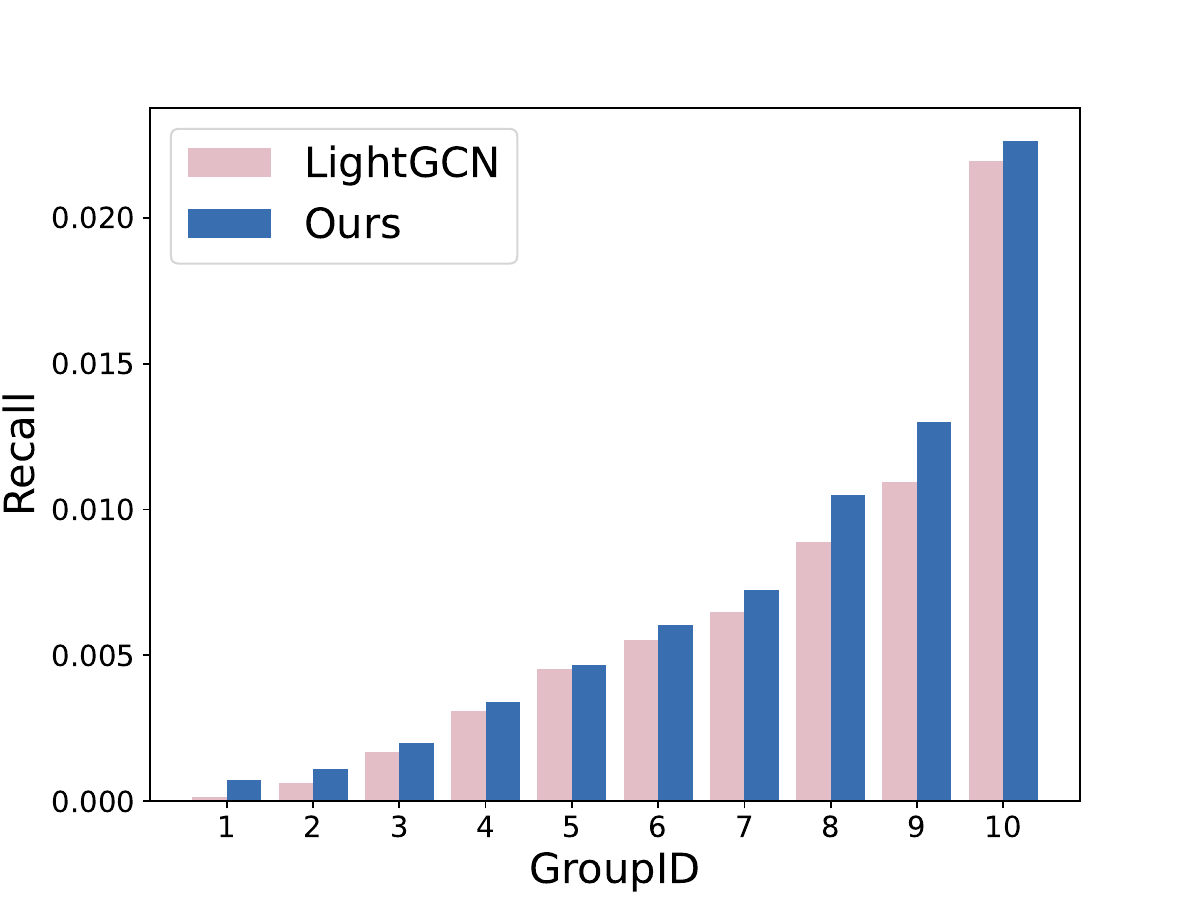} \label{fig:g1}} 
\subfloat[Amazon-book]{\includegraphics[width=0.49\columnwidth]{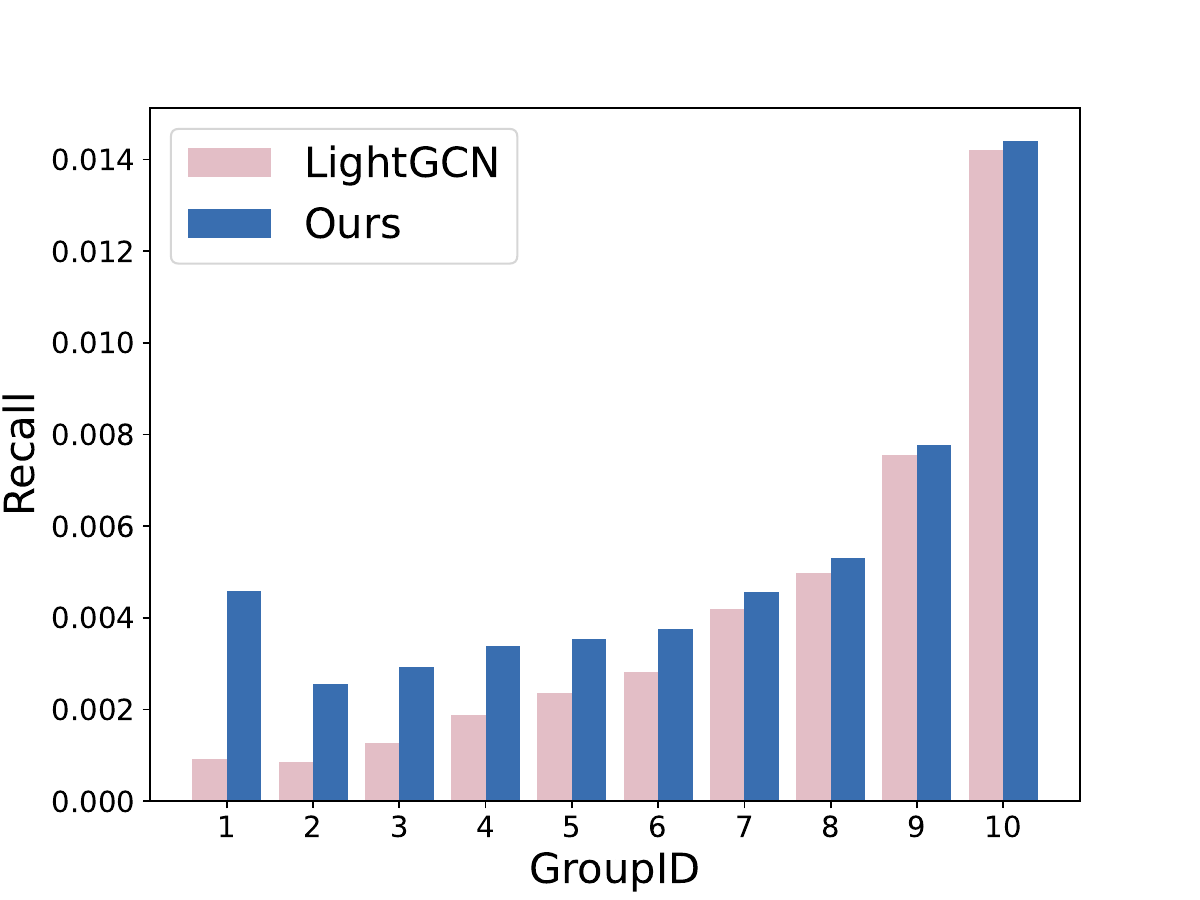} \label{fig:g2}}
% \subfloat[Movielens-1m]{\includegraphics[width=0.33\textwidth]{figs/3c.pdf} \label{fig:g3}} \\   
   \caption{Performance comparison over item groups.}
   \vspace{-0.4cm}
\label{fig:groups}
\end{figure}

\subsection{Recommendation Performance Comparison}

We evaluate the model from two aspects, overall recommendation performance and cold-start recommendation performance.

\subsubsection{Overall Performance.}

The details of the comparison are listed in Table \ref{tab:mainrr}. The following points are observed:

\noindent $\bullet$ Our UCC outperforms previous models significantly on all datasets. On the Yelp dataset, our method outperforming recall of LightGCN by 11.6\%. Compared to the state-of-the-art graph based method SGL, our method also obtains a higher recall of SGL by 6\%. The improvement is consistent for Amazon-Book. This demonstrates the effectiveness of our uncertainty-aware consistency learning. 

\noindent $\bullet$ Our method improves Recall@20 significantly more on the Amazon-Book dataset - 27.6\% compared to LightGCN and 9.6\% higher than SGL. The reason is that the interactions for cold items in amazon-book are more biased and sparse. This validates that our method targets at the cold-start problem and well alleviates the restriction of different distribution between cold items and warm items.

\subsubsection{Cold-Start Performance.} 
We further compare our methods with LightGCN on cold and warm items separately to illustrate its effectiveness for cold-start recommendation. We split items into ten groups based on the popularity, meanwhile keeping the total number of interactions of each group the same. We treat the last two groups as cold-start items. The larger the GroupID is, the warmer the items are. We evaluate our method on different groups and plot the Recall@20 in Figure \ref{fig:groups}. From the results, we notice that our method outperforms LightGCN on all groups, no matter whether the items are warm or cold. In comparison, many current methods improve the performance of cold items at the cost of warm item accuracy. Also, our method improves the recall of LightGCN more prominently on cold items. On the Yelp dataset, our method obtains nearly 7 times compared to recall of LightGCN. The most significant improvement comes from the groupid 1 items of Amazon-Book. In this group, the recall of amazon-book obtains is 0.0009. In contrast, the recall of our method obtains 0.0045, improved by 400\%. This strongly demonstrates the effectiveness of our method for cold-start recommendation and the problem of the seesaw phenomenon.

\subsection{Ablation Study of UCC}

% \begin{figure} %[h]
%   % \centering
% \subfloat[Generated interaction num]{\includegraphics[height=0.40\columnwidth]{figs/5a.pdf} \label{fig:apl1}}
% \subfloat[Performance comparison]{\includegraphics[height=0.40\columnwidth]{figs/5b.pdf} \label{fig:apl2}}
%    \caption{Analysis of generated interactions. (a) the number of generated interactions is adaptive on warm and cold items, (b) item-side generated interactions produce higher Recall@20.}
%    \label{fig:apl}
% \end{figure}

In this section, we perform ablation studies to evaluate each component of UCC, which consists of two parts: Uncertainty-aware Interaction Generation and Teacher-Student Consistency Learning.

\subsubsection{Uncertainty-aware interaction generation}
We calculate the average number of generated interactions for different item groups. We find the number of generated interactions is adaptive for different items. For cold items, the average number of produced interactions is about 3.5, which is about 1.0 for warm items in comparison. Since the generated low-uncertainty interactions contribute more for cold items, the uncertainty guided strategy well alleviating the different distribution between warm items and cold items.
We also notice that the simple usage of item-side generated interactions helps improves the performance. The performance increase indicates that our item-side adaptive generated interactions can indeed alleviate the cold-start problem. User-side generated interactions, in comparison, cannot improve the recommendation ability of the model - even lowering the Recall@20 from 0.0673 to 0.0671. This is because user-side generated interactions are biased and exacerbate the distribution between cold and warm items. This demonstrates the superiority of our uncertainty-aware interaction generation.

\subsubsection{Teacher-Student Consistency Learning}
% We first illustrate the necessity of strong and weak augmentation in consistency learning. Without any augmentation strategy, our teacher-student learning approach helps improve the Recall@20 from 0.0628 to 0.0651. The recall of this approach is also higher than the simple usage of generated interactions (0.0614). This further validates the effectiveness of our designed teacher-student framework. After weak augmentation based consistency regularization is introduced, the Recall@20 is improved to 0.071. The performance increase comes from better generalization and robustness ability brought by the consistency regularization. The strong augmentation further helps the Recall@20 improve to 0.0713, since noise in generated interactions provides stronger consistency regularization.

% \begin{table}
% \centering
% \caption{Study of Weak-Strong augmentation.}
% \begin{tabular}{cc|cc}
% \hline 
% Teacher & Student & Recall & NDCG\\
% \hline 
% w/o & - & 0.0628 & 0.0515  \\
% w/o & w/o & 0.0651 & 0.0530  \\
% weak & weak & 0.0710 & 0.0586 \\
% weak & strong  & 0.0713 & 0.0587 \\
% \hline
% \end{tabular}
% \label{tab:strongweak}
% \end{table}

\begin{figure} %[h]
\includegraphics[width=0.88\columnwidth]{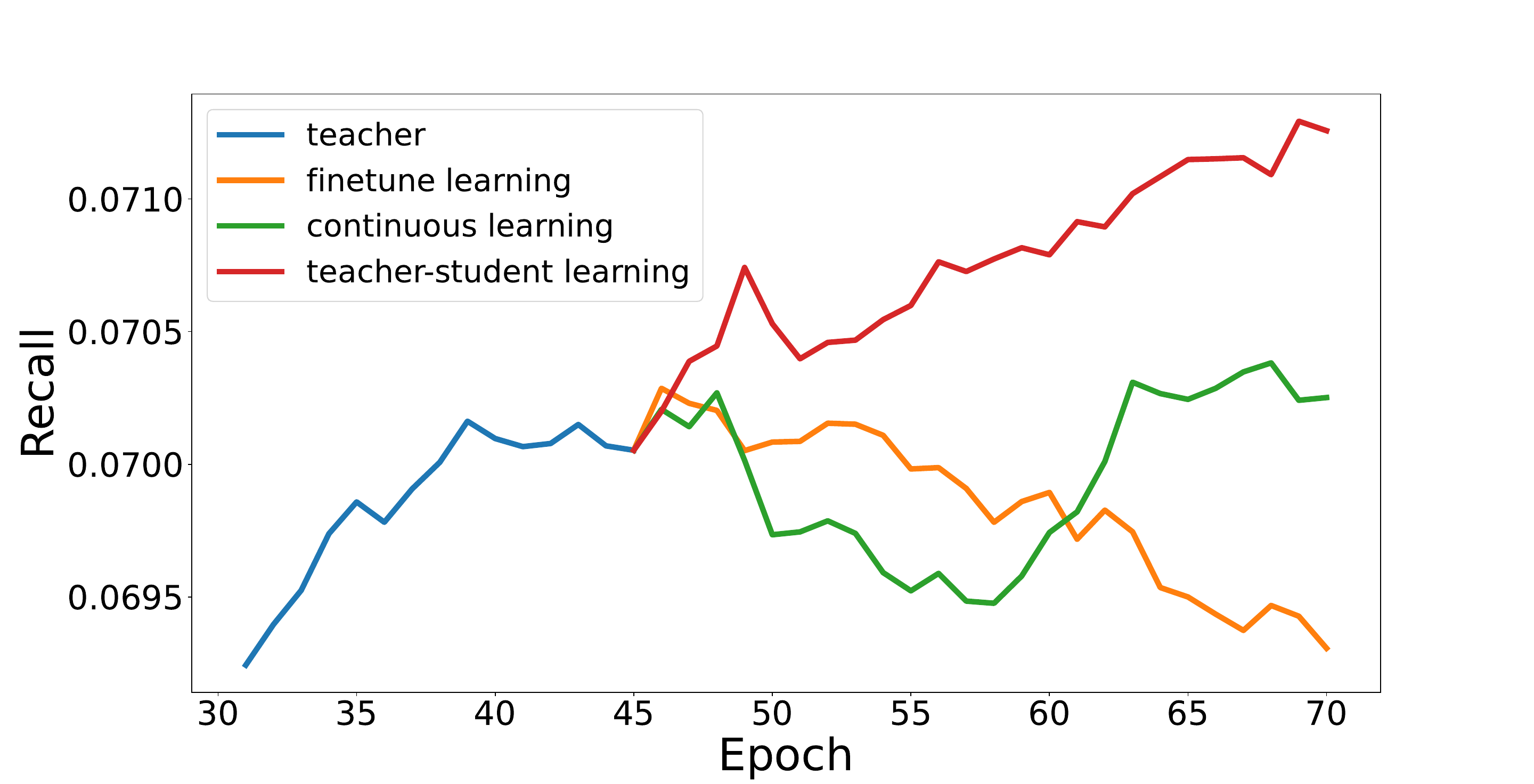}
\caption{Study of teacher-student learning.}
\vspace{-0.4cm}
\label{fig:ts}
\end{figure}

We show the effectiveness of our teacher-student learning approach. We design two comparing experiments: 1) Finetune learning:  We train the student representations with generated interactions directly from the teacher model embeddings without accumulating teacher representations. 2) Continuous learning: We train the student embeddings with accumulating with teacher embeddings without using generated interactions. We plot the results in Figure \ref{fig:ts}.
We observe that our teacher-student learning approach obtains a higher performance. The superior comes from two aspects. Compared with finetune learning, our student model maintains more information from the teacher model. The information within the teacher embeddings alleviates the negative effect of noise from generated interactions. Compared with continuous learning, we provide more interactions for cold items, thus mitigating the cold-start problem. Through the proposed teacher-student learning framework, similar distributions of information and data are accumulated together, thus helping obtain a better recommendation performance.

% \input{4.related.tex}
% vspace{-10cm}
\section{Conclusion}\label{sec::conclusion}
In this paper, we identify that the main limitation comes from the interaction-distribution difference between cold and warm items. To address it, we propose an uncertainty-aware consistency learning framework. 
The extraordinary performance on benchmark datasets and its easy-to-use property make our framework practical in both academia and industry.

%Our work presents an initial attempt to exploit pseudo labels for graph based long-tail recommendation and opens up new research possibilities. In the future, we will explore more effective ways to utilize pseudo labels and alleviate the long-tail problem. The main point is how to provide more reliable and realistic pseudo labels for recommendation models and how to avoid the negative effect of pseudo labels. Another possible direction is to promote information communication between teacher and student models. We hope our model is beneficial for improving the generalization and robustness of recommender models.

\section{ACKNOWLEDGEMENT}
This work was supported by the National Key Research and Development Program of China under 2022YFB3104702, the National Natural Science Foundation of China under 62272262, 61972223, U1936217, and U20B2060, the Fellowship of China Postdoctoral Science Foundation under 2021TQ0027 and 2022M710006.

\bibliographystyle{ACM-Reference-Format}
\balance
\bibliography{sample-base}

\end{document}